%% file: Template.tex
\documentclass{article}
\usepackage{spconf, amsmath, graphicx}
\usepackage{float}
\usepackage{subfigure}
\usepackage{cite}
\usepackage[urlcolor=blue]{hyperref}

\makeatletter
\newcommand*\bigcdot{\mathpalette\bigcdot@{.5}}
\newcommand*\bigcdot@[2]{\mathbin{\vcenter{\hbox{\scalebox{#2}{$\m@th#1\bullet$}}}}}
\makeatother

\title{improving language model-based zero-shot text-to-speech synthesis with multi-scale acoustic prompts}

\name{
\begin{tabular}{@{}c@{}}
Shun Lei$^{1\ddagger}$\thanks{$^{\ddagger}$Work done when the first author was intern at Skywork AI PTE. LTD.}, Yixuan Zhou$^{1\dagger}$, Liyang Chen$^{1}$, Dan Luo$^{1}$, Zhiyong Wu$^{1,3*}$\thanks{$^{\dagger}$ Equal contribution. $^{*}$ Corresponding author.}, Xixin Wu$^{3*}$, \\ 
\textit{Shiyin Kang$^{2*}$, Tao Jiang$^2$, Yahui Zhou$^2$, Yuxing Han$^{1}$, Helen Meng$^3$}
\end{tabular}
}

\address{
    $^1$ Shenzhen International Graduate School, Tsinghua University, Shenzhen\\ 
    $^2$ Skywork AI PTE. LTD., Beijing $^3$ The Chinese University of Hong Kong, Hong Kong SAR\\
    \small{
        \{leis21, yx-zhou23\}$@$mails.tsinghua.edu.cn, 
        zywu$@$sz.tsinghua.edu.cn,
        wuxx$@$se.cuhk.edu.hk, 
        shiyin.kang$@$kunlun-inc.com
    }
}
\begin{document}
\ninept
%
\maketitle

\input{abstract}
\input{introduction}

\input{methodology}

\input{experiments}
\input{conclusions}

\vfill\pagebreak

\bibliographystyle{IEEEbib}
\bibliography{strings,refs}
\end{document}

%% file: abstract.tex
\begin{abstract}
Zero-shot text-to-speech (TTS) synthesis aims to clone any unseen speaker’s voice without adaptation parameters.
By quantizing speech waveform into discrete acoustic tokens and modeling these tokens with the language model, recent language model-based TTS models show zero-shot speaker adaptation capabilities with only a 3-second acoustic prompt of an unseen speaker.  
However, they are limited by the length of the acoustic prompt, which makes it difficult to clone personal speaking style. 
In this paper, we propose a novel zero-shot TTS model with the multi-scale acoustic prompts based on a language model.
A speaker-aware text encoder is proposed to learn the personal speaking style at the phoneme-level from the style prompt consisting of multiple sentences.
Following that, a VALL-E based acoustic decoder is utilized to model the timbre from the timbre prompt at the frame-level and generate speech.
The experimental results show that our proposed method outperforms baselines in terms of naturalness and speaker similarity, and can achieve better performance by scaling out to a longer style prompt\footnote{Speech sample: \href{https://thuhcsi.github.io/icassp2024-msvalle}{https://thuhcsi.github.io/icassp2024-msvalle}}.
\end{abstract}

\begin{keywords}
text-to-speech, zero-shot, multi-scale acoustic prompts, speaker adaptation, language model
\end{keywords}

%% file: introduction.tex
\vspace{-0.1cm}
\section{Introduction}
\vspace{-0.1cm}
Text-to-speech (TTS) aims to generate natural and intelligible speech from text. 
With the development of deep learning, neural network based TTS models can already synthesize high-quality speech for single \cite{tacotron2, fastspeech2} or multiple speakers \cite{chen2020multispeech, vits}.
However, these models still require a sufficient amount of clean speech data for new speakers, which hinders the development of speech synthesis technology for many personalized applications.
Therefore, adapting TTS models for any speaker with as few data as possible, while achieving high speaker similarity and speech naturalness, has attracted increasing interest in academia and industry \cite{survey}.

One of the general approaches is fine-tuning a well-trained multi-speaker TTS model with a few adaptation data to support new speakers.
Some studies devote effort to fine-tune the whole TTS model \cite{chen2018sample, Kons2019}, and other recent methods seek to reduce the adaptation parameters by fine-tuning only a part of the model \cite{moss2020boffin}, or only the speaker embedding \cite{chen2020adaspeech}.
However, the adaptation performance of these methods relies heavily on the quality and quantity of the data available for the target speaker. 

To deal with this deficiency, some works conduct zero-shot adaptation, which leverages only a few seconds of speech to clone an unseen speaker's voice without fine-tuning the model.
In \cite{jia2018transfer, cooper2020zero, casanova2022yourtts}, a speaker encoder is utilized to extract global speaker embeddings from the given reference speech, which allows the TTS model to clone the overall timbre of the reference speech.
Considering that it is difficult to describe the personal characteristics of speakers with a single speaker embedding, \cite{choi20c_interspeech, zhou22d_interspeech} propose to extract fine-grained speaker embeddings to improve the quality of synthesized speech.
Motivated by advancements in natural language generation models, recent speech generation systems \cite{borsos2023audiolm,wang2023neural,kharitonov2023speak} introduce the idea of utilizing neural audio codec \cite{defossez2022high,zeghidour2021soundstream} to quantize speech waveform into discrete tokens and leveraging prompting-based language model (e.g., GPT-3 \cite{brown2020language}) to predict these tokens.
These language model-based TTS systems can be trained on large, diverse and low-quality multi-speaker speech datasets to improve generalization performance.
With these approaches, the models are capable of cloning the speaker's timbre with only a 3-second acoustic prompt.

However, the above language model-based zero-shot TTS methods only consider the acoustic prompt at frame-level, leading to two major limitations.
First, the speaker characteristics of a person include not only the timbre but also personal speaking style, which consists of various elements such as prosody, accents and pronunciation habits.
While considering the acoustic prompt at frame-level has shown the great power of timbre clone, it has been proven that phoneme-level representations are more suited for generating personal speaking style \cite{10202208,jiang2023mega}.
Second, limited by the structure of the decoder-only language model, these works only support a short acoustic prompt because the frame-level acoustic token sequence is too long (a 10s speech usually contains thousands of tokens).
It is difficult to use the limited information contained in the short acoustic prompt to accurately clone the speaker characteristics of the target speaker, leading to poor naturalness and similarity of speaking style.
In addition, current language model-based methods have no ability to utilize multiple reference samples to enhance the quality of zero-shot TTS even though several utterances of the target speaker are available during inference in many real-world scenarios.

To further improve speaker similarity for language model-based zero-shot TTS synthesis, we propose to utilize multi-scale acoustic prompts to capture both the timbre and personal speaking style of the target speaker. 
Our model contains a speaker-aware text encoder, which utilizes a reference attention module to model the personal speaking style at phoneme-level from the style prompt consisting of multiple utterances and an acoustic decoder, which preserves a specified timbre by considering timbre prompt at frame-level based on the neural codec language model (called VALL-E).
The model allows scaling out to an arbitrary length of style prompt to describe detailed speaker characteristics.
Both subjective and objective evaluations show that our proposed method outperforms state-of-the-art language model-based zero-shot TTS model \cite{wang2023neural} and other baselines in terms of naturalness and speaker similarity. The performance is also improved with an increasing number of sentences used in the style prompt during inference.

%% file: methodology.tex
\vspace{-0.1cm}
\section{methodology}
\vspace{-0.1cm}

\begin{figure}[!tb]
	\centering
	\includegraphics[width=0.6\linewidth]{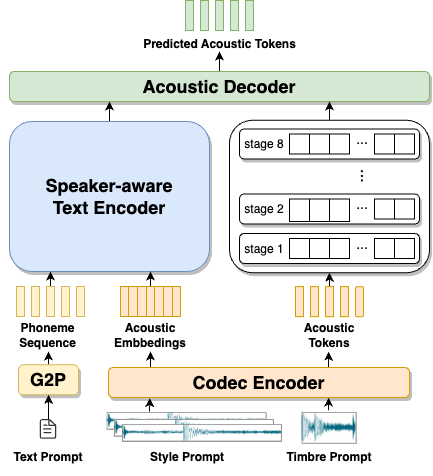}
	\caption{The overall architecture of the proposed model}
	\label{fig:model}
\end{figure}
The architecture of our proposed model is illustrated in Fig.\ref{fig:model}.
It consists of two major parts: a speaker-aware text encoder and an acoustic decoder based on VALL-E.
In this paper, we follow Naturalspeech 2 \cite{shen2023naturalspeech} and VALL-E \cite{wang2023neural} to leverage neural audio codec models to represent style prompt and timbre prompt in continuous acoustic embeddings and discrete acoustic tokens, respectively.
The speaker-aware text encoder is used to extract phoneme-level personal speaking style information from the style prompt and fuse it into encoded phoneme embeddings by a reference attention module to obtain speaker-aware text embeddings.
Then the outputs of the encoder are fed into the acoustic decoder along with the acoustic tokens of the timbre prompt to generate speech with the same timbre as the timbre prompt.
The details of each component are as follows.

\vspace{-0.1cm}
\subsection{Speaker-aware Text Encoder}
\begin{figure}[!tb]
	\centering
    \includegraphics[height=0.6\linewidth]{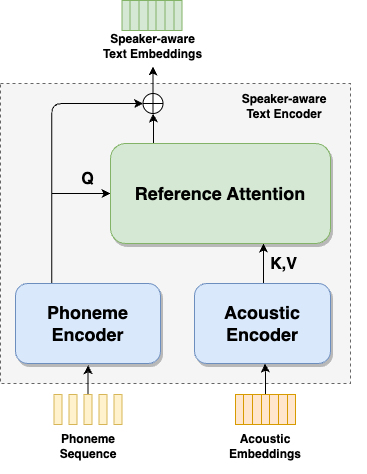}
	\caption{The structure of the speaker-aware text encoder}
	\label{fig:encoder}
\end{figure}

The speaker-aware text encoder is specifically designed to extract and model personal speaking style at phoneme-level from an arbitrary-length style prompt and fuses the text-side content information with the speech-side style information to obtain the speaker-aware text embeddings.
The architecture of the encoder is illustrated in Fig.\ref{fig:encoder}, which comprises a phoneme encoder, an acoustic encoder and a reference attention module.

On the text side, to derive better text representation as decoder input, we introduce a phoneme encoder to encode the phoneme sequence.
We use the ransformer block, which is a stack of self-attention layer and 1D-convolution as in Fastspeech 2 \cite{fastspeech2}, as the basic structure for the encoder.
The input texts are converted into a sequence of phonemes by the grapheme-to-phoneme module and then passed to the phoneme encoder to obtain phoneme embeddings.

On the speech side, to make use of arbitrary-length speech prompts, previous approaches attempt to encode the speaker characteristics into a global-level vector \cite{cooper2020zero, casanova2022yourtts}.
As a result, the local fine-grained variations in speaking style are ignored.
Different from this way, we use an acoustic encoder to derive the local speaking style embeddings from the style prompt instead of a single vector.
All the utterances of the target speaker are firstly concatenated to form the style prompt, and then passed to a well-trained neural audio codec to convert speech waveform into continuous acoustic embeddings instead of discrete tokens to preserve as much personal style information as possible in the speech.
Then converted acoustic embeddings are then passed to the acoustic encoder, which is made up of a stack of 8 1D-convolution layers.
In addition, in order to regulate the temporal granularity of the extracted style representations closer to human vocal perception, the filter strides of convolution layers are set as [2,1,2,1,2,1,2,1] for 16 times downsampling (about 0.2s).
After that, the temporal granularity of the style embeddings is properly reformed to a quasi-phoneme level inspired by \cite{multiscale}.

To make better use of style embeddings extracted from the style prompt, a reference attention module is introduced to obtain the appropriate phoneme-level semantic-related personal speaking style.
We adopt scaled dot-product attention as the reference attention module.
The phoneme embeddings are regarded as the query, while all the style embeddings extracted from the style prompt are regarded as both the key and the value.
The relevance between them is used to guide the selection of the personal speaking style for each input phoneme.
Finally, the reference attention module outputs an aligned sequence with the same length as the phoneme embeddings and adds it to the phoneme embeddings to form the speaker-aware text embeddings.

\vspace{-0.1cm}
\subsection{Acoustic Decoder}
In order to model speaker characteristics of a person, it is necessary to clone the timbre in addition to mimic the speaker's speaking style.
Inspired by the success of language models in zero-shot TTS, our proposed method adopts a modified VALL-E \cite{wang2023neural} as the acoustic decoder to generate speech with the same timbre as the 3-second timbre prompt.
As illustrated in Fig.\ref{fig:decoder}, the decoder is made up of an acoustic embedding, an autoregressive (AR) transformer decoder and a non-autoregressive (NAR) transformer decoder.

The timbre prompt is first passed to a well-trained neural audio codec, and the output of the residual vector quantizer in the codec is considered as discrete prompt acoustic tokens.
These tokens consist of 8 layers which are then embedded through eight separate acoustic embedding layers.
The AR transformer decoder is utilized to generate the first layer of acoustic tokens required to synthesis personalized speech conditioned on the speaker-aware text embeddings.
Meanwhile, the first layer of acoustic tokens of the timbre prompt is used as the prefix in AR decoding.
The NAR transformer decoder is then used to generate acoustic tokens of the other seven layers in sequence.
To predict the acoustic tokens of the $i$-th layer, the transformer input is the concatenation of the speaker-aware text embeddings, the summation of the embedded acoustic tokens of timbre prompt from layer 1 to layer $i$ and the summation of the embedded predicted acoustic tokens from layer 1 to layer $i-1$.
In the end, the first layer of acoustic tokens predicted by the AR transformer decoder and the remaining layers of acoustic tokens predicted by the NAR transformer decoder are concatenated to form the predicted acoustic tokens.

\vspace{-0.1cm}
\subsection{Training Strategy and Inference Procedure}
During the training stage, for each training sample, we randomly select 5 to 10 reference utterances spoken by the same speaker as the sample to form the style prompt.
For different training epochs, different style prompts are randomly selected for the same training sample for data augmentation.
Different from VALL-E which trains two models separately, our proposed method trains the whole end-to-end system jointly with the cross-entropy loss.
Training loss is a linear combination of an AR transformer decoder loss and an NAR transformer decoder loss.
In the AR transformer decoder, we do not explicitly select an utterance as the timbre prompt in training, which means all acoustic tokens of the first layer are predicted with the teacher-forcing technique. 
For the NAR transformer decoder, in each training step, we randomly sample a training stage $i\in [2,8]$ and randomly select a certain length of target speech prefixes as the timbre prompt.
The model is trained to maximize the probability of the acoustic tokens in the $i$-th layer.

During the inference stage, we design acoustic prompts and inference as follows.
To generate given content for unseen speakers, the model is given a text sentence, any number of speeches from the target speaker as the style prompt, a short segment of speech from the target speaker as the timbre prompt and its corresponding transcription.
We prepend the transcription of the timbre prompt to the given text sentence as the text prompt.
With the text prompt, the style prompt and the timbre prompt, our proposed method generates the acoustic tokens for the given text cloning the speaker characteristics of the target speaker.

%% file: experiments.tex
\vspace{-0.1cm}
\section{Experiments}
\vspace{-0.1cm}

\subsection{Training Setup}
All the models are trained on LibriTTS \cite{libritts}, which is an open-source multi-speaker transcribed English speech dataset.
Its training set contains approximately 580 hours of recording spoken by 2,306 speakers.
To evaluate the zero-shot adaptation capability for unseen speakers, 128 speakers from two subsets of the LibriTTS dataset (test-clean and dev-clean) are selected as the test set, resulting in 8,078 utterances in total.
A pre-trained neural audio codec model, EnCodec\footnote{Implemented based on: \href{https://github.com/facebookresearch/encodec}{https://github.com/facebookresearch/encodec}} \cite{defossez2022high}, is utilized as the codec model to encode the raw waveform with 24kHz sampling rate and reconstruct the waveform based on the predicted acoustic tokens.

In our implementation, the phoneme encoder, AR transformer decoder and NAR transformer decoder all consist of 6 layers of transformer blocks.
Compared to the two modules in the original VALL-E which both consist of 12 layers of transformer blocks, our proposed model has less parameters.
We train all the models for 300K iterations on 4 NVIDIA A100 GPUs, with a batch size of 8 samples on each GPU.
The Adam optimizer is adopted with $\beta_1=0.9$, $\beta_2=0.98$ and follow the same learning rate schedule in \cite{wang2023neural}.

\vspace{-0.1cm}
\subsection{Compared Methods}

\begin{figure}[!tb]
	\centering
	\includegraphics[width=1\linewidth]{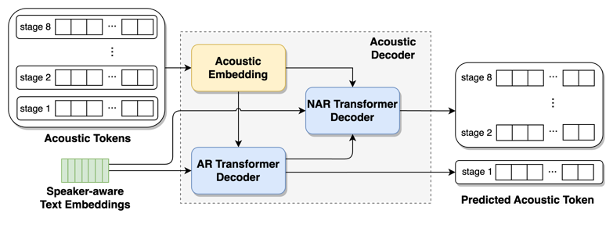}
	\caption{The structure of the acoustic decoder}
	\label{fig:decoder}
\end{figure}

To demonstrate the performance of our proposed method, we compare the following five models for zero-shot TTS synthesis. These models are also implemented based on VALL-E\footnote{Implemented based on: \href{https://github.com/lifeiteng/vall-e}{https://github.com/lifeiteng/vall-e}\label{valle}}.

\textbf{VALL-E}
An open-source implementation\textsuperscript{\ref{valle}} of VALL-E \cite{wang2023neural}, which considers only a 3-second timbre prompt at the frame-level.

\textbf{Proposed}
The proposed model, which considers both a 3-second timbre prompt and a style prompt consisting of ten sentences.

\textbf{Proposed-3s}
To ensure a fair comparison, we build this baseline model, which shares the same structure and parameters as the proposed model, but only uses a 3-second speech as both timbre prompt and style prompt.

\textbf{Base-S}
The style prompt-only baseline model, which shares the same TTS backbone and style prompt as the proposed model, but excludes the timbre prompt.

\textbf{Base-T}
The timbre prompt-only baseline model, where the style prompt is removed. That is, this model used only a 3-second speech as the timbre prompt.

For each sample synthesis, we randomly choose other utterances of the same speaker as the timbre prompt and the style prompt.

\vspace{-0.1cm}
\subsection{Subjective Evaluation}
We conduct two mean opinion score (MOS) tests to measure the zero-shot capability of different models: 1) Naturalness MOS (N-MOS): evaluate the naturalness and prosody of the synthesized speech; 2) Similarity MOS (S-MOS): evaluate the speaker similarity between the synthesized speech and the ground-truth speech.
We randomly choose 20 samples from different speakers in the test set for subjective evaluation.
To exclude other interference factors, we keep the text content and prompt speech consistent among different models.
A group of 25 listening subjects are recruited to rate the given speeches on a scale from 1 to 5 with 1 point interval. 

\begin{table*}[h]
  \caption{The objective and subjective comparisons for zero-shot text-to-speech synthesis. We evaluate the naturalness and speaker similarity of different models with with 95\% confidence intervals.}
  \label{tab:mos}
  \centering
  \begin{tabular}{c|cc|cc} 
    \hline
    \textbf{Models} & \multicolumn{2}{c}{\textbf{Subjective}} & \multicolumn{2}{c}{\textbf{Objective}}  \\
    \textbf{} & \textbf{N-MOS} ($\uparrow$) & \textbf{S-MOS} ($\uparrow$) & \textbf{SECS} ($\uparrow$)  & \textbf{MCD} ($\downarrow$)  \\
    \hline
    Ground Truth & $4.23\pm0.066$ & - & - & - ~~~               \\
    \hline
    VALL-E & $3.48\pm0.059$& $3.532\pm0.060$ & $0.771$ & $8.075$ ~~~  \\
    Base-S & $3.456\pm0.055$ & $3.500\pm0.056$& $0.727$& $7.792$  ~~~  \\
    Base-T & $3.654\pm0.062$ & $3.646\pm0.060$& $0.764$& $8.047$ ~~~ \\
    \hline
    Proposed & $\mathbf{3.886\pm0.063}$ & $\mathbf{3.870\pm0.062}$ & $\mathbf{0.798}$& $\mathbf{7.715}$~~~ \\
    Proposed-3s & $3.778\pm0.063$ & $3.692\pm0.062$ & ${0.779}$& ${7.765}$~~~ \\
    \hline
  \end{tabular}
  \vspace{-0.6cm}
\end{table*}
\begin{table}[h]
  \caption{The objective evaluation results of VALL-E and proposed method when using different length of prompt in inference. The average duration of sentences is about 6 seconds.}
  \label{tab:moreexp}
  \centering
  \begin{tabular}{c|cc} 
    \hline
    \textbf{Models} & \textbf{SECS} ($\uparrow$)  & \textbf{MCD} ($\downarrow$)  \\
    \hline
    VALL-E w/ 3s& $0.771$ & $8.075$  ~~~               \\
    VALL-E w/ 6s& $0.774$ & $8.177$  ~~~               \\
    \hline
    Proposed w/ 1 sent (3s) & $0.779$ & $7.765$  ~~~               \\
    Proposed w/ 5 sent (30s)& $0.795$& $7.743$ ~~~  \\
    Proposed w/ 10 sent (1min)& $0.798$ & $7.715$ ~~~  \\
    Proposed w/ 20 sent (2min)& $0.798$ & $7.702$ ~~~ \\
    \hline
  \end{tabular}
  \vspace{-0.6cm}
\end{table}

As shown in Table \ref{tab:mos}, our proposed method achieves the best N-MOS of $3.886$ and S-MOS of $3.870$, which outperforms VALL-E greatly in both two aspects.
Compared with the VALL-E and the Base-T model, which both used only 3 seconds of speech as the acoustic prompt, the Proposed-3s model achieves better performance, especially in naturalness by a gap of over $0.12$. 
This demonstrates that considering the same acoustic prompt at the phoneme-level is really helpful for learning the personal speaking style of the target speaker and improving the naturalness and speaker similarity of the synthesized speech without introducing additional inputs.
Our proposed model also makes further improvement compared with the Proposed-3s model by increasing the number of sentences in the style prompt, 
indicating the ability of  enhancing  the quality of zero-shot TTS by utilizing more reference speeches from the target speaker .
This ability is not available in previous language model-based TTS models, and it allows our proposed method to have a higher performance upper bound compared to VALL-E.
Our proposed model also achieves superior performance than not only Base-S that just considers the style prompt, but also Base-T that solitarily considers the timbre prompt.
It demonstrates that modeling the speaker characteristics from different scales can improve the naturalness and speaker similarity of the synthesized speech.
In addition, it is observed that although Base-S also uses ten sentences as the style prompt, it achieved the lowest score in both two evaluations.
A possible reason is that removing the timbre prompt as the prefix in the decoder affects the stability of the decoding, resulting in some synthesized speeches are poor in intelligibility.
Moreover, we added a comparison with YourTTS \cite{casanova2022yourtts}, through the ABX preferene test.
Our proposed model showed a $57.3\%$ preference rate over YourTTS's $33.6\%$, demonstrating its effectiveness.

\vspace{-0.1cm}
\subsection{Objective Evaluation}

To measure the naturalness and speaker similarity of synthesized speech objectively, we calculate mel-cepstrum distortion (MCD) and Speaker Encoder Cosine Similarity (SECS) as the metrics following \cite{casanova2022yourtts, choi20c_interspeech}.
Since the lengths of the predicted and ground-truth speech may be different, we ﬁrst apply dynamic time warping (DTW) to derive the alignment relationships between the two mel-spectrograms. Then, we compute the minimum MCD by aligning the two mel-spectrograms.
For the speaker similarity, we use the speaker encoder of the Resemblyzer package\footnote{Implemented based on: \href{https://github.com/resemble-ai/Resemblyzer}{https://github.com/resemble-ai/Resemblyzer}} to compute the SECS between the ground-truth speech and synthesized speech.
The value ranges from 0 to 1, where a large value indicates a higher similarity.

The evaluation results of different models on the test set are shown in Table \ref{tab:mos}.
It is observed that our proposed model outperforms the baselines in all objective evaluation metrics.
The results indicate that our proposed model can improve the 
quality 
and speaker similarity of synthesized speech.

\vspace{-0.1cm}
\subsection{Investigation}
To investigate the impact of the prompt with different lengths, we adjust the length of the acoustic prompt and style prompt for VALL-E and the proposed model, respectively.
For VALL-E, constrained by the structure of the decoder-only language model, we randomly select two utterances of 3s/6s as the prompts for each speaker. 
We also evaluate our proposed model with various numbers of speech as the style prompt, including 1 sentence, 5 sentences, 10 sentences and 20 sentences.
The timbre prompt is fixed to 3-second speech as mentioned above.
In particular, when the style prompt consists of only one sentence, the proposed model only uses a 3-second speech as both the timbre prompt and style prompt.
We evaluate these models with the two objective metrics as described before.

Table \ref{tab:moreexp} shows the performance comparison among the different lengths of the acoustic prompt.
It is observed that the VALL-E with an acoustic prompt of 6 seconds speech gets the SECS result close to the proposed method with only one sentence as style prompt, but there is a significant gap with the proposed in MCD.
It demonstrates that modeling the personal speaking style of the target speaker at phoneme-level helps generate speech that is close to the ground-truth.
By comparing different lengths of the style prompt, we can see our proposed model is able to generate more similar speech when the number of sentences in style prompt increases.

%% file: conclusions.tex
\vspace{-0.1cm}
\section{Conclusions}
\vspace{-0.1cm}
In this paper, we propose a language model-based zero-shot TTS model to utilize multi-scale acoustic prompts to capture both the timbre and personal speaking style of the target speaker.
A speaker-aware text encoder is utilized to model the speaking style at phoneme-level from arbitrary-length style prompt.
A language model-based acoustic decoder is used to preserve a specified timbre by considering the timbre prompt at frame-level.
Experimental results demonstrate that our proposed approach could significantly improve the naturalness and speaker similarity of the synthesized speech, even when only using 3-second speech as both style prompt and timbre prompt.
In addition, our proposed model can enhance the quality of zero-shot TTS by increasing the number of sentences in style prompt, when there are multiple sentences of the target speaker are available during inference.

\vspace{+0.3cm}
\textbf{Acknowledgement}: This work is supported by National Natural Science Foundation of China (62076144), National Social Science Foundation of China (13\&ZD189), Shenzhen Science and Technology Program (WDZC20220816140515001, JCYJ202208181010140\\30) and Shenzhen Key Laboratory of next generation interactive media innovative technology (ZDSYS20210623092001004).